\documentclass{amsproc}

\theoremstyle{definition}

\theoremstyle{remark}

\numberwithin{equation}{section}

\catcode`\@=11
\def\@citex[#1]#2{%
\if@filesw \immediate \write \@auxout {\string \citation {#2}}\fi
\@tempcntb\m@ne \let\@h@ld\relax \def\@citea{}%
\@cite{%
  \@for \@citeb:=#2\do {%
    \@ifundefined {b@\@citeb}%
      {\@h@ld\@citea\@tempcntb\m@ne{\bf ?}%
      \@warning {Citation `\@citeb ' on page \thepage \space undefined}}%
      {\@tempcnta\@tempcntb \advance\@tempcnta\@ne%
      \@tempcntb\number\csname b@\@citeb \endcsname \relax%
      \ifnum\@tempcnta=\@tempcntb 
        \ifx\@h@ld\relax%
          \edef \@h@ld{\@citea\csname b@\@citeb\endcsname}%
        \else%
          \edef\@h@ld{\ifmmode{-}\else--\fi\csname b@\@citeb\endcsname}%
        \fi%
      \else
        \@h@ld\@citea\csname b@\@citeb \endcsname%
        \let\@h@ld\relax%
      \fi}%
    \def\@citea{,\penalty\@highpenalty\,}%
  }\@h@ld
}{#1}}

\def\@citeb#1#2{{[#1]\if@tempswa , #2\fi}}
\def\@citeu#1#2{{$^{#1}$\if@tempswa , #2\fi }}
\def\@citep#1#2{{#1\if@tempswa , #2\fi}}

\newcommand{\beqa}{\begin{eqnarray}}
\newcommand{\eeqa}{\end{eqnarray}}

\newcommand{\dd}{{\rm d}}

\newcommand{\Z}{{\mathbb Z}}

\newcommand{\R}{{\mathbb R}}
\newcommand{\C}{{\mathbb C}}
\newcommand{\PP}{{\mathbb P}}
\newcommand{\e}{\,{\rm e}}
\newcommand{\CP}{{\C\PP}}

\newcommand{\bPhi}{\overline{\Phi}}
\newcommand{\ttheta}{\widetilde{\theta}}
\newcommand{\bD}{\overline{D}}
\newcommand{\tW}{\widetilde{W}}
\def\cal{\mathcal}
\newcommand{\bQ}{\overline{Q}}

\begin{document}



\title{Mirror Symmetry And Some Applications}

\author{Kentaro Hori}
\curraddr{Jefferson Physical Laboratory, 
Harvard University,
Cambridge, MA 02138, U.S.A.}
\email{hori@infeld.harvard.edu}

\date{June 5, 2001}


\begin{abstract}
We report on recent progress in understanding mirror symmetrey.
Some of more recent generalizations and applications are also presented.\\
{\it A contribution to the Proceedings of
``Strings 2001'' at Mumbai, India.}
\end{abstract}

\maketitle

\section{Introduction}

Mirror symmetry is originally conjectured as equivalence of
$(2,2)$ superconformal field theories in $1+1$ dimensions, under which
vector and axial $U(1)$ R-symmetries are exchanged \cite{Dixon}.
Since the discovery of mirror pairs of Calabi-Yau manifolds \cite{GP},
it has played important roles in exploring quantum geometry
of string theory.
Mirror symmetry is also interesting from mathematical point of view:
Symplectic geometric aspects of one theory is reflected in complex analytic
aspects of the mirror, and difficult problems in the former are
solved using classical techniques of the latter. 
The conjecture has also been extended
to $(2,2)$ theories without conformal invariance \cite{FI}.

Mirror symmetry is nothing but T-duality for sigma models on complex tori,
and there is also a proposal that mirror symmetry between
Calabi-Yau three-folds is T-duality on a supersymmetric $T^3$ fibration
\cite{SYZ}.
In this contribution,
we report on the progress \cite{HV} in understanding
mirror symmetry from an exact field theory analysis,
including a proof in some class of theories.
Mirror symmetry is indeed T-duality applied to gauge systems
that flow to non-linear sigma models.
This, however, turns sigma models into Landau-Ginzburg models,
where the LG potential for the dual fields
is generated by the vortex-anti-vortex gas
of the original gauge system
(as in Polyakov's story of confinement
in $2+1$ dimensional QED that includes monopole-instantons
\cite{Polyakov}).
In the conformal cases, we need one more step to make
LG models into sigma models on the mirror manifolds.
In the non-conformal cases, sigma model realization of the mirror
is forbidden on the symmetry ground.

We will also present some recent generalizations and applications
of \cite{HV},
including the ${\cal N}=2$ duality of 2d Black Hole
and Liouville theory \cite{HK},
D-branes \cite{HIV,H,HN,AV}, and relation to
mirror symmetry in $2+1$ dimensional gauge theories \cite{AHKT}.


\section{A Proof Of Mirror Symmetry}\label{sec:proof}

The class of models we consider here are non-linear sigma models on
toric manifolds, including asymptotically free sigma models
on compact manifolds of positive first Chern
classes (such as $\CP^{N-1}$), as well as those on
non-compact Calabi-Yau manifolds (such as the resolved conifold)
that flow to conformal field theories.
Toric sigma models are realized as low energy theories
of certain gauge systems called
linear sigma models \cite{oldLSM,phases}.

Consider a $U(1)^k$ gauge theory with $N$ chiral multiplets $\Phi_i$
of charge $Q_i^a$ ($i\!=\!1,...,N$, $a\!=\!1,...,k$).
The classical Lagrangian is given by
\begin{equation}
L=\int\dd^4\theta \left[\,\sum_{i=1}^N\bPhi_i\e^{Q_i\cdot V}\Phi_i
-\sum_{a=1}^k{1\over 2e_a^2}|\Sigma_a|^2\,\right]
+{\rm Re}\int \dd^2\ttheta \,\sum_{a=1}^k(-t^a\Sigma_a),
\label{LSM}
\end{equation}
where $Q_i\cdot V:=\sum_{a=1}^kQ_i^aV_a$ and $\Sigma_a:=\bD_+D_-V_a$
(fieldstrength).
The theory is classically
parametrized by the dimensionful gauge coupling constants $e_a$
 and dimensionless FI-Theta parameters $t^a=r^a-i\theta^a$.
It is super-renormalizable with respect to $e_a$
but $r^a$ has to be renormalized as
$r^a(\mu')=r^a(\mu)+b_1^a\log(\mu'/\mu)$,
$b_1^a:=\sum_{i=1}^NQ_i^a$
(causing a dimensional transmutation if $b_1^a\ne 0$).
The system has an exact vector $U(1)$ R-symmetry but
the axial $U(1)$ R-rotation $\e^{iF_A\beta}$ shifts
$\theta^a$ by $-2b_1^a\beta$.

The system has a flavor symmetry group $U(1)^{N-k}$ rotating
$\Phi_i$ with charges
complementary to $Q_i^a$.
For a suitable choice of $r^a$,
the space of classical vacua is a toric manifold $X$ of dimension $N-k$
where the $U(1)^{N-k}$ flavor group action
determines the structure of the torus fibration.
The metric on $X$ is obtained by the standard K\"ahler reduction.
For example, for $U(1)$ gauge theory with two charge 1 chiral fields
the classical moduli space is $X=\CP^1$ with the round (Fubiny-Study) metric. 
At energies much smaller than $e\sqrt{r}$,
the theory reduces to the supersymmetric non-linear sigma model on $X$.
The $\theta^a$ angles determine the B-field.
$X$ is a (non-compact) Calabi-Yau manifold if and only if all $b_1^a=0$.

Let us now dualize the phases of $\Phi_i$.
We first consider the simplest case of $U(1)$ gauge theory with a single
chiral multiplet $\Phi$ of charge $1$.
The standard dualization procedure \cite{BuRV} shows that the dual of a
chiral superfield is a twisted chiral superfield, and if $\Phi$ is charged
the dual is neutral but is coupled to the gauge
multiplet as the dynamical FI-Theta parameter.
Thus, the dual theory is in terms of a twisted chiral superfield $Y$
(with periodicity $2\pi i$) which is linearly coupled to
the gauge multiplet in the twisted superpotential
as $\Sigma Y$.
The original and the dual fields are related by
\begin{equation}
\bPhi\e^V\Phi={\rm Re}\,Y.
\end{equation}
One can show that $Y$ shifts as
$Y\to Y-2i\beta$ under axial R-rotation, reflecting the axial anomaly.
Also, the renormalization of the FI parameter forces
the renormalization of ${\rm Re}Y$ as $Y(\mu')=Y(\mu)+\log(\mu'/\mu)$.
By this effect, the semi-classical K\"ahler metric for $Y$
becomes close to the flat one, even though we are dualizing along the
circle of varying radius.

This is not the whole story.
The original gauge system includes vortex configurations
where $\Phi$ vanishes at some points around which the phase of $\Phi$ has
non-trivial winding numbers.
Since T-duality exchanges winding number and momentum,
the vortex configuration is mapped to the insertion of momentum-creation
operator which is of the type $\e^{-i{\rm Im}Y}$.
There are vortex configurations that preserve half of
the $(2,2)$ supersymmetry, so that the twisted superpotential
can be generated. By the standard instanton computation,
one can show that $\e^{-Y}$, the supersymmetric completion
of the momentum-creation operator, is indeed generated by such a
supersymmetric vortex of unit charge.
Note that the anomalous axial R-transformation of $Y$ is such that
$e^{-Y}$ has the right axial R-charge.
Together with the tree-level term $-t\Sigma$ and the term
$\Sigma Y$ that appears at the dualization,
we obtain the twisted superpotential of the dual theory,
\begin{equation}
\tW=\Sigma(Y-t)+\e^{-Y}.\label{Wexact}
\end{equation}
This is the exact twisted superpotential,
as can be shown by noting that
no other expression preserves symmetry, holomorphy and conditions
on asymptotic behaviour (that any correction has to be small
for large $t$, $Y$ and small $\Sigma$).

Let us come back the general case. By dualization on the phase,
each $\Phi_i$ turns into a twisted chiral
superfield $Y_i$ with period $2\pi i$.
We claim that the exact twisted superpotential of the dual theory
is simply
\begin{equation}
\tW=\sum_{a=1}^k\Sigma_a\left(\sum_{i=1}^NQ_i^aY_i-t^a\right)
+\sum_{i=1}^N\e^{-Y_i}.\label{Wexact2}
\end{equation}
The $\Sigma_aQ_i^aY_i$ terms appear at the dualization level, while
the exponential terms are the ones generated by non-perturbative
effects.
The easiest way to show this is to gauge the flavor symmetry group
$U(1)^{N-k}$ so that the total gauge group is $U(1)^N$. 
For a suitable choice of the gauge coupling, the system
is the sum of $N$ decoupled systems, each being a copy of 
the $k=N=1$ system considered above.
Then, the dual twisted superpotential is the sum over $i\!=\!1,...,N$
of $\Sigma_i(Y_i-t_i)+\e^{-Y_i}$.
This is true for any value of the gauge couplings since
the twisted superpotential can depend only on twisted chiral parameters.
We recover the original system
by taking the weak coupling limit of the flavor group $U(1)^{N-k}$.
This limit simply constrains the gauge fields and the FI-Theta parameters
as $\Sigma_i=\sum_{a=1}^k Q_i^a\Sigma_a$,
$\sum_{i=1}^NQ_i^at_i=t^a$.
This yields (\ref{Wexact2}).
(There is a room of shifting $\Sigma_i$ by an $i$ dependent constant.
This corresponds to the twisted mass deformation \cite{HH}
of the linear sigma model
that reduces to the potential deformation in the sense of \cite{AF}.)

The original linear sigma model reduces to the non-linear sigma model on
the toric manifold $X$ by sending $e_a\to\infty$.
In the dual theory, this requires
integrating out the heavy gauge multiplets $\Sigma_a$.
By the $\Sigma_a$-linear terms in (\ref{Wexact2}), this yields the
constraints
\begin{equation}
\quad \sum_{i=1}^NQ_i^aY_i=t^a\quad(\mbox{mod $2\pi i\Z$}).
\label{constraint}
\end{equation}
This is the equation defining a $(\C^{\times})^{N-k}$
subspace of $(\C^{\times})^N=\{(\e^{-Y_i})\}$.
Thus, we obtain the LG model on this $(\C^{\times})^{N-k}$
with the superpotential
\begin{equation}
\tW=\sum_{i=1}^N\e^{-Y_i}.\label{Wmir}
\end{equation}
This is the mirror of the non-linear sigma model on the toric
manifold $X$.

We have derived mirror symmetry between sigma models and LG models.
Can we also derive more standard mirror symmetry between
sigma models of different manifolds?
First of all, if $X$ is not Calabi-Yau, axial $U(1)$ R-symmetry is
anomalously broken and therefore the mirror cannot be just a sigma model
(that would have an exact vector $U(1)$ R-symmetry).
For Calabi-Yau sigma models this obstruction is absent.
Here we present a way to relate our mirror LG model
to a sigma model via a simple change of variables.
We present it only in the example of the resolved conifold
since the generalization is obvious.
The linear sigma model for the resolved conifold is the $U(1)$ gauge theory
with four chiral multiplets with charges $1,1,-1,-1$.
The mirror LG model has superpotential $W=\sum_{i=1}^4\e^{-Y_i}$
for variables with the constraint $Y_1+Y_2-Y_3-Y_4=t$.
The constraint can be solved by
$Y_1=Y_0$, $Y_3=Y_0+\Theta_1$, $Y_4=Y_0+\Theta_2$ and
$Y_2=t+Y_0+\Theta_1+\Theta_2$ and the superpotential has a factorized form
$W=\e^{-Y_0}(1+\e^{-t-\Theta_1-\Theta_2}+\e^{-\Theta_1}+\e^{-\Theta_2})$.
Now, we can add to the system a massive LG model $W_1=UV$
without affecting the IR limit. By the change of variables
$U\to \e^{-Y_0}u$, $V\to v$, the total superpotential again factorizes
$W+W_1=\e^{-Y_0}(1+\e^{-t-\Theta_1-\Theta_2}+\e^{-\Theta_1}+\e^{-\Theta_2}
+uv)$.
The holomorphic measure $\dd Y_0\dd\Theta_1\dd\Theta_2\dd U\dd V$
is expressed as $\e^{-Y_0}\dd Y_0\dd\Theta_1\dd\Theta_2\dd u\dd v$, and we
see that $\e^{-Y_0}$ becomes a $\C$-variable rather than
a $\C^{\times}$-variable. Integration over $\e^{-Y_0}$ imposes a constraint
\begin{equation}
1+\e^{-t-\Theta_1-\Theta_2}+\e^{-\Theta_1}+\e^{-\Theta_2}+uv=0.
\end{equation}
In this way, the mirror LG model is related to a sigma model
on a manifold defined by this equation.
As one can see, this argument only shows
the ``relation'' as a partial equivalence where only
holomorphic information is conserved.
This is in contrast with the mirror symmetry between the toric sigma
model and the LG model, which is an exact equivalence.


\section{The Fermionic 2d Black Hole And ${\cal N}=2$ Liouville Theory}
\label{sec:2dBH}

What we have done above is essentially T-duality with respect to
the torus fibration determined by 
the $U(1)^{N-k}$ global symmetry.
This group action is not free and the circle fibre for $\arg \Phi_i$
shrinks to zero size at the locus $|\Phi_i|=0$.
One would na\"\i vely expect that
the dual torus blows up at the same locus.
However, what we have seen shows that a superpotential term $\e^{-Y_i}$
is generated in the dual theory.
This diverges toward the locus $|\Phi_i|=0$, and breaks the
rotational invariance of the dual theory, accounting for the loss of
winding number in the sigma model due to the degeneration of the circle.

A while ago, Fateev, Zamolodchikov and Zamolodchikov
\cite{FZZ} conjectured a duality between the two-dimensional black hole
background and a potential theory
called {\it sine-Liouville theory}.\footnote{This
was recently used in the Matrix Model formulation of
string theory in the black hole background \cite{KKK}
(c.f. the contribution by V.~A.~Kazakov).
More recent studies are in \cite{reFZZ}.}
The former has a semi-infinite cigar geometry
which asymptotes to a flat cylinder with a linear dilaton.
The latter is a theory of a cylinder variable with
(Liouville $\times$ cosine) potential.
This also has an asymptotic region with a linear dilaton
where the potential is exponentially small.
Comparison of the asymptotic regions
suggests that the two theories are related by T-duality.
The degeneration of the circle in the 2d black hole
corresponds to the growing sine-Liouville potential.
This is strongly reminiscent of our mirror symmetry.

In \cite{HK} a proof is given for the supersymmetric version of the
FZZ duality \cite{GK}
(c.f. \cite{MV,ES}).
It is between the 2d black hole background for fermionic strings,
$SL(2,\R)_{k+2}$ mod $U(1)$ Kazama-Suzuki super-coset model,
and ${\cal N}=2$ Liouville theory,
the LG model of a periodic chiral superfield $Y\equiv Y+2\pi i$
with the superpotential $W=\e^{-Y}$ and the K\"ahler potential
$K=|Y|^2/2k$.
The two theories
are weakly coupled in the opposite regimes
$k\gg 1$ and $k\ll 1$.

The essential part of the proof is to show that the SCFT of the 2d black
hole arizes as the infra-red fixed point of the following
$U(1)$ gauge theory. It is the theory with two chiral superfields
$\Phi$ and $P\equiv P+2\pi i$ which transforms as
$\Phi\to \e^{i\alpha}\Phi$
and $P+i\alpha$ under the $U(1)$ gauge transformation,
with the Lagrangian given by
\begin{equation}
L=\int\dd^4\theta\left[\,\bPhi\e^V\Phi+{k\over 4}(P+\overline{P}+V)^2
-{1\over 2e^2}|\Sigma|^2\,\right].
\end{equation}
The classical vacuum manifold is a semi-infinite cigar, but
the metric is not precisely that of the 2d black hole and the
dilaton is not linear in the asymptotic region.
However, a one-loop renormalization group analysis
shows that it flows to the 2d black hole background,
including the linear dilaton.
This analysis is valid at large $k$.
Next, following \cite{SW},
one computes the central charge of the infra-red fixed point,
by identifying the ${\cal N}=2$ superconformal algebra in the ring of
left-chiral operators.
This is possible when one can identify the right-moving R-current.
The axial $U(1)$ R-symmetry is anomalous in the present system,
but one can use the field
$P$ to modify the current in a gauge invariant way so that it is conserved.
By reality of the currents in the asymptotic resion,
one can also fix the ambiguity due to the
presence of the global symmetry.
Thus, one can completely identify the ${\cal N}=2$ algebra.
This shows that the central charge of the IR fixed point is
\begin{equation}
c=3\Bigl(\,1+{2\over k}\,\Bigr),
\end{equation}
which is the correct value for the super-coset model.
The modified parts of the ${\cal N}=2$ currents are linear in $P$
and this exhibits the linear dilaton
in the asymptotic region (with the correct slope).
Finally, one excludes the possibility that the theory flows not to the
super-coset itself but to some other nearby fixed point
with the same central charge, symmetries, and asymptotic behaviour.
This is done by showing that the super-coset has no supersymmetric, parity
invariant marginal operator that is small at the asymptotic region.

The rest is a straightforward generalization of the argument in the
previous section.
Dualization of the phase of $\Phi$ and the imaginary part of $P$
turns them into twisted chiral superfields $Y$ and $Y_P$
(both period $2\pi i$).
The exact superpotential of the dual system is
\begin{equation}
\tW=\Sigma(Y+Y_P)+\e^{-Y}.
\end{equation}
No non-perturbative superpotential is generated for $Y_P$ since
there is no $P$-vortex.
The K\"ahler potential is
\begin{equation}
K=-{1\over 2e^2}|\Sigma|^2-{1\over 2k}|Y_P|^2+\cdots,\label{kahler}
\end{equation}
where $\cdots$ are small in the asymptotic region.
In the limit $e\to\infty$, it is appropriate to integrate out $\Sigma$
and we have the constraint $Y+Y_P=0$.
In fact, one can use the rigidity of the super-coset
to show that the terms $\cdots$ in (\ref{kahler}) vanishes.
Thus, we conclude that the dual theory flows to the ${\cal N}=2$
Liouville theory with the K\"ahler potential $K=-|Y|^2/2k$.
This completes the proof of the claimed equivalence.

This story has a generalization which can be presented
as a deformation of the standard linear sigma model (\ref{LSM}).
We gauge the $U(1)^{N-k}$ flavor group, acting on $\Phi_i$
with charge $R_{iq}$ ($q=1,\ldots,N-k$), and introduce
for each $U(1)$ a chiral superfield $P_{q}$ transforming
inhomogeneously.
The Lagrangian of the system reads
\beqa
\,\,\,\,L\!\!\!&=&\!\!\!\int\dd^4\theta\Biggl[\,
\sum_{i=1}^N\bPhi_i \e^{Q_i\cdot V+R_i\cdot V'}\Phi_i
-\sum_{a=1}^{k}
{1\over 2e_{a}^2}|\Sigma_{a}|^2\,\Biggr]
+{\rm Re}\int \dd^2\widetilde{\theta}\sum_{a=1}^k(-t^a\Sigma_a)
\\
&&~~~~~~~~~~~~~~~~~~
+\int\dd^4\theta\Biggl[\,
\sum_{q=1}^{N-k}{k_{q}\over 4}(P_{q}+\overline{P}_{q}+V'_{q})^2
-\sum_{q=1}^{N-k}{1\over 2e_{q}^2}|\Sigma'_{q}|^2
~\Biggr]\nonumber
\eeqa
where $R_i\cdot V'=\sum_{q=1}^{N-k}R_{iq}V'_{q}$.
The vacuum manifold $X'$ is
again a toric manifold with the same K\"ahler class as $X$,
but with a different metric.
Deep in the interior of the base of the torus fibration,
the sizes of the fibers
are constants $\sim\sqrt{k_{q}}$.
Thus $X'$ is a ``squashed'' version of the toric manifold.
In the limit $k_{q}\to\infty$,
the $P$-$\Sigma'$ pairs decouple, and
we recover the sigma-model on the ``round toric manifold'' $X$.
When all $b_1^a=0$,
the theory is expected to flow to a non-trivial superconformal field
theory.
The ${\cal N}\!=\!2$ currents can be uniquely identified in the left-chiral
ring if there are asymptotic regions, and they include linear terms
in $P_{q}$ proportional to $b_{q}\!:=\!\sum_{i=1}^{N}R_{iq}$,
implying a linear dilaton if $b_q\ne 0$.
The central charge is found as
\begin{equation}
c=3\Bigl(\,N-k+\sum_{q=1}^{N-k}{2b_{q}^2\over k_{q}}\,\Bigr).
\end{equation}
In the ``round limit'' $k_{q}\to\infty$, we recover
$c/3\to N-k=\dim X$.

The dual theory is found as before.
In the limit $e_a,e_q\to\infty$, we obtain a LG model on
the $(\C^{\times})^{N-k}$ of (\ref{constraint}) with
the following K\"ahler and superpotentials
\begin{equation}
K=-{1\over 2}\sum_{i,j=1}^N
g_{ij}\overline{Y}_iY_j+\cdots,\,\,\,\,\,\,
\widetilde{W}=
\sum_{i=1}^N\e^{-Y_i},\,\,\,
\label{KW}
\end{equation}
where $g_{ij}:=\sum_{q=1}^{N-k}R_{iq}R_{jq}/k_{q}$.
This is the mirror of the sigma-model on the squashed toric manifold $X'$.
Note that the K\"ahler potential depends on the
squashing parameters $k_{q}$ and tends to vanish in the
``round limit'' $k_{q}\to \infty$.
For $X=\CP^1$, $X'$ is sausage-shaped, and
we find that the mirror of the supersymmetric sausage
model is the ${\cal N}=2$ sine-Gordon model with a finite K\"ahler potential.
This equivalence has been conjectured by Fendley and
Intriligator.

\section{D-Branes}\label{sec:DO}

\newcommand{\Bgamma}{\mbox{\large $\gamma$}}

It is a natural idea to ask how mirror symmetry acts on
D-branes and how it can be used.
D-branes can be regarded as boundary conditions or boundary interactions
on the worldsheet of an open string.
We will restrict ourselves to those preserving a half of
the $(2,2)$ worldsheet supersymmetry. (Here we do not consider space-time
supersymmetry.) There are essentially two kinds of such D-branes
\cite{OOY}:
A-branes preserving the combinations $Q_A=\bQ_++Q_-$ and
$Q_A^{\dag}=Q_++\bQ_-$;
B-branes preserving $Q_B=\bQ_++\bQ_-$ and $Q_B^{\dag}=Q_++Q_-$.
Since mirror symmetry exchanges $Q_-$ and $\bQ_-$,
A-branes and B-branes are exchanged under mirror symmetry.

Let us consider the non-linear sigma model on a K\"ahler manifold $X$.
$X$ can be considered as a complex manifold or as a symplectic manifold
(with respect to the K\"ahler form $\omega$).
A D-brane wrapped on a cycle $\Bgamma$ of $X$ and supporting
a $U(1)$ gauge field $A$ is an A-brane if $\Bgamma$
is a Lagrangian submanifold ($\omega|_{\gamma}=0$) and $A$ is flat ($F_A=0$),
while it is a B-brane if $\Bgamma$ is a complex submanifold of $X$ and
$A$ is holomorphic ($F_A^{(2,0)}=0$).
If we consider a LG model for the superpotential $W$ (a holomorphic
function of $X$), there is a further condition that
the $W$-image of $\Bgamma$ is a straightline parallel to the real axis
for A-branes and $W$ is a constant on $\Bgamma$ for B-branes.
A-branes and B-branes
are objects of interest from the point of view of symplectic
geometry and complex analytic geometry, respectively.
They are exchanged under mirror symmetry,
as anticipated by M.~Kontsevich \cite{Kon}.

The theory of an open string stretched between two A-branes (or two B-branes)
can be considered as a supersymmetric quantum mechanics
with infinitely many degrees of freedom.
We present the analysis of supersymmetric ground states
in the four cases \cite{HIV,H}:
{\bf (i)} A-branes in massive LG models,
{\bf (ii)} space filling B-branes in sigma models,
{\bf (iii)} B-type D0-branes in massive LG models,
{\bf (iv)} A-branes wrapped on the fibres of a torus fibration $X$.
For {\bf (i)} and {\bf (ii)},
the quantum mechanics has a $U(1)$ R-symmetry: $U(1)_A$ for {\bf (i)}
and $U(1)_V$ for {\bf (ii)} (for LG we assume $X$ is non-compact
Calabi-Yau).
{\bf (iii)} and {\bf (iv)} admit topological twisting:
B-twist for {\bf (iii)} and A-twist for {\bf (iv)}.\\[0.1cm]
{\bf (i)} Let $p_a$ and $p_b$ be critical points of the superpotential
$W$. Gradient flows of ${\rm Re}\, W$ originating from $p_a$ and $p_b$ sweep
Lagrangian submanifolds $\Bgamma_a$ and $\Bgamma_b$ whose $W$-images are
parallel straight lines emanating from the critical values
$W_a, W_b$.
We consider D-branes wrapped on $\Bgamma_a$ and $\Bgamma_b$.
Classical supersymmetric open string configurations
are gradient flows of
$-{\rm Im}\,W$ from a point in $\Bgamma_a$ to
a point in $\Bgamma_b$.
If ${\rm Im}\,W_a<{\rm Im}\, W_b$ there is of course no such flow
and hence no supersymmetric ground states.
If ${\rm Im}\,W_a>{\rm Im}\, W_b$, the number of such flows,
counted with an appropriate sign, is the intersection number
$S_{ab}=\Bgamma_a\cap \Bgamma_b'$ where $\Bgamma_b'$ is the deformation
of $\Bgamma_b$ so that the $W$-image
is rotated at $W_b$ with an infinitesimal positive angle.
($S_{ab}$ is the number of BPS solitons
from $p_a$ to $p_b$ when there is no critical value between
$W$ images of $\Bgamma_a$ and $\Bgamma_b$.)
Quantum mechanically, the paths with opposite signs are lifted
by instanton effects and $S_{ab}$ is indeed the number of
supersymmetric ground states.
One can also quantize the space of paths obeying the D-brane
boundary condition with a Morse function determined by
$W$ and the K\"ahler form $\omega$. The space of supersymmetric ground
states is identified as the cohomology
${\rm HF}_{W}^{\bullet}(\Bgamma_a,\Bgamma_b)$
of the Morse-Witten complex \cite{SQM}.
(This is studied also by Y.-G.~Oh \cite{Oh}.)\\[0.1cm]
{\bf (ii)} Consider D-branes supporting holomorphic vector
bundles $E_a$ and $E_b$ over $X$.
The zero mode sector of the open string Hilbert space
is identified as the space
$\Omega^{0,\bullet}(X,E_a^{*}\otimes E_b)$ where the supercharge
$Q_B$ acts as the Dolbeault operator.
Thus, the space of supersymmetric ground states
in the zero mode approximation is the Dolbeault cohomology
or ${\rm Ext}^{\bullet}(E_a,E_b)$.
In the full theory, there is nothing more than this, but
some pairs of neighboring R-charges could be lifted.
The latter does not happen if ${\rm Ext}^{p}(E_a,E_b)$ is non-zero
only for even $p$ (or odd $p$).
An example is a pair $E_a,E_b$ from an {\it exceptional collection}
\cite{rudakov},
which is an ordered set of bundles $\{E_i\}$
where ${\rm Ext}^p(E_i,E_i)=\delta_{p,0}\C$
while for $i<j$ ${\rm Ext}^{\bullet}(E_j,E_i)=0$ but
${\rm Ext}^{p}(E_i,E_j)$ can be non-zero
only for one value of $p$.\\[0.1cm]
{\bf (iii)} Consider D0-branes located at two critical points $p_a$
and $p_b$ of the superpotential. 
There is no supersymmetric ground states if $p_a\ne p_b$.
For $p_a=p_b$, there are $2^n$ supersymmetric ground states, half bosonic
and half fermionic, where $n=\dim X$.
In the quadratic approximation of the superpotential, this
is shown by explicitly quantizing the system.
More precisely, one can use the one-to-one correspondence
(based on B-twist) between the supersymmetric ground states
and the boundary chiral ring elements. 
The boundary chiral ring is the algebra of $n$ anti-commuting
elements $\overline{\theta}^1,\ldots,\overline{\theta}^n$,
and this gives the number $2^n=2^{n-1}+2^{n-1}$.\\[0.1cm]
{\bf (iv)}
Let $\Bgamma_a$ and $\Bgamma_b$ be fibres of a torus fibration $X$,
and let $A_a$ and $A_b$ be flat $U(1)$ connections on them.
We consider D-branes $(\Bgamma_a,A_a)$ and $(\Bgamma_b,A_b)$.
The open string Witten index is the intersection number of
$\Bgamma_a$ and $\Bgamma_b$ which is zero.
For the case $\Bgamma_a=\Bgamma_b$,
the zero mode sector is identified as the space
$\Omega^{\bullet}(X)$ where the supercharge
$Q_A$ acts as the operator $d+A_b-A_a$.
There is no supersymmetric ground states if $A_a\not\equiv A_b$
while if $A_a\equiv A_b$ there are $2^n$ of them,
half bosonic and half fermionic.
This is actually exact since the boundary
chiral ring is $H^{\bullet}(T^n)=\C^{2^n}$.

An important characteristic of a D-brane $a$ is the overlap of the boundary
state $\langle a|$ and the RR ground states $|i\rangle$
of the bulk (closed string) theory,
$\Pi^a_i=\langle a|i\rangle$.
In Type II or Type I string theory, it measures the
coupling of the D-branes to RR fields, or the
D-brane charges (times tension).
Let us consider the case where $a$ is an A-brane.
$\Pi^a_i$ is non-zero only if $|i\rangle$ has zero axial R-charge.
If the bulk theory is B-twistable,
it is convenient to use as $|i\rangle$ the ones corresponding
to the chiral ring elements $\phi_i$.
Then, $\Pi^a_i$ is independent of the twisted chiral parameter
but depends on the chiral parameters as a parallel
section with respect to the so called improved connection \cite{OOY,HIV}.
If the bulk theory is A-twistable,
$\Pi_i^a$ is the topological Disk amplitude if we use as $|i\rangle$
the ones corresponding to twisted chiral ring elements.
(All can be repeated for B-branes under the replacement
axial$\to$vector,
B-twist$\leftrightarrow$A-twist,
chiral$\leftrightarrow$twisted chiral.)
For Calabi-Yau sigma models the overlap for A-branes
is the classical period integral
$\Pi_i^a=\int_{\gamma_a}\Omega_i$ \cite{OOY}.
For LG A-branes in {\bf (i)}, it is a weighted period
integral $\Pi_i^a=\int_{\gamma_a}\phi_i\e^{-W}\Omega$ \cite{HIV}.
For LG B-branes in {\bf (iii)},
the overlap is all vanishing, $\Pi_i^a=0$ \cite{H}.

D-branes in the ${\cal N}=2$ minimal model can be
studied by considering massive deformations of the LG model $W=X^{k+2}$
\cite{HIV,H}.
A-branes are oriented wedges which asymptote to the $k+2$
inverse imange lines of $W\in \R_{\geq 0}$.
(A similar geometric picture emerges from the study of the $SU(2)_{k}$
mod $U(1)$ super-coset realization \cite{MMS,Scho}.)
The intersection numbers $\Bgamma_a\cap\Bgamma_b'$ provide a
geometric realization of the Fusion coefficients, and
the integral $\Pi_j^a\sim \int_{\gamma_a}X^{2j}\e^{-X^{k+2}}\dd X$ reproduces
the known values for the Cardy states \cite{Cardy} written in terms of
the modular S-matrix.
B-branes in a deformed model 
are $(k+1)$ D0-branes at the critical points,
all of which have vanishing RR-overlaps $\Pi_j^a=0$.
This slightly disagrees with the result
at the conformal point \cite{MMS}:
there are $2[k/2]+1$ of them with $\Pi_j^a=0$
and, for even $k$, there are extra ones
with non-trivial RR-overlaps $\Pi_{j}^a\ne 0$
(on the $j=k/4$ state with zero R-charge).
It would be interesting to understand how the extra brane (dis)appears
after massive deformations.

Below, we see how the class of D-branes
in {\bf (i)}-{\bf (iv)} are related under mirror symmetry discussed in
\S\ref{sec:proof}.

\subsection{B-Branes on $X$
$\longleftrightarrow$ A-Branes of LG}

To be specific, we consider the case $X=\CP^2$ where the mirror is the
LG model on $(\C^{\times})^2$ with the superpotential
$W=\e^{-Y_1}+\e^{-Y_2}+\e^{-t+Y_1+Y_2}$.
Corresponding to the three vacua of the $\CP^2$ sigma model,
$W$ has three critical points $\e^{-Y_1}=\e^{-Y_2}=\e^{-t/3},
\e^{-(t+2\pi i)/3},\e^{-(t-2\pi i)/3}$ which we denote
by $p_0,p_1,p_{-1}$ respectively.
Let us first consider the B-brane
supporting the trivial line bundle ${\cal O}(0)$.
By lifting the Neumann boundary condition to the linear
sigma model and following the dualization procedure, we find that
the mirror of ${\cal O}(0)$ is the Lagrangian submanifold $\Bgamma_0$
of type {\bf (i)} which emanates from $p_0$
in the positive real direction in its $W$-image.
Let us now perform an axial R-rotation which shifts the $\theta$ angle
by $2\pi$. The shift turns ${\cal O}(0)$
into ${\cal O}(-1)$. In the mirror side, the same R-rotation rotates
$W(\Bgamma_0)$ around $W=0$ by $120^{\circ}$ and it now emanates from
$p_{-1}$ and extend in the radial direction $W\in\e^{2\pi i/3}\R_{>0}$.
This preserves the supercharge
$\e^{-\pi i/3}\bQ_++\e^{\pi i/3}Q_-$ which is the R-rotation of $Q_A$.
To find the configuration that preserves $Q_A$ itself, we need to rotate
it back by $-120^{\circ}$ around $W(p_{-1})$ so that it emanates from
$p_{-1}$ in the positive real direction.
We call it $\Bgamma_{-1}$. This is the mirror of ${\cal O}(-1)$
preserving $Q_B$.
Repeating the same procedure for the opposite R-rotation,
we find that the mirror of ${\cal O}(1)$ is the cycle
$\Bgamma_1$ which emanates from $p_1$
in the positive real direction in the $W$-image.

In fact $\{{\cal O}(-1),{\cal O}(0),{\cal O}(1)\}$ is an exceptional
collection, and therefore the space of supersymmetric ground states
of an open string stretched between these branes is given by
\begin{equation}
\begin{array}{l}
{\rm Ext}^{p}({\cal O}(-1),{\cal O}(0))=\delta_{p,0}\C^3\\
{\rm Ext}^{p}({\cal O}(0),{\cal O}(1))=\delta_{p,0}\C^3\\
{\rm Ext}^{p}({\cal O}(-1),{\cal O}(1))=\delta_{p,0}\C^6
\end{array}
\label{ext}
\end{equation}
and ${\rm Ext}^{p}({\cal O}(j),{\cal O}(i))=0$ for $i<j$.
On the other hand, since
${\rm Im}\,W(p_j)<{\rm Im}\,W(p_i)$, $i<j$,
there is no supersymmetric ground states for open string from $\Bgamma_j$ to
$\Bgamma_i$ if $i<j$, in agreement with the last statement.
Also, since there are no critical values between
the $W$-images of $\Bgamma_{-1}$ and $\Bgamma_0$, the number of
supersymmetric ground states is the number of BPS solitons interpolating
$p_{-1}$ and $p_0$. This is $3$ (which can be checked explicitly
or by using the fact that the solitons are identified as the
fundamental electrons $\Phi_i$ in the linear sigma model \cite{oldLSM}).
This is in agreement with the first line of (\ref{ext}).
The second line is similar.
To check the third line, we need to compute the intersection number
of $\Bgamma_{-1}$ and $\Bgamma_1'$. This differs from the soliton number
$-3$ since there is a critical value $W(p_0)$ between their $W$-images.
The difference is, by Picard-Lefchetz theory, given by 
$(\Bgamma_{-1}\cap\Bgamma_0')\cdot (\Bgamma_0\cap\Bgamma_1')=9$.
Thus we find $\Bgamma_{-1}\cap\Bgamma_1'=-3+9=6$, in agreement with
the third line.
All these agreements are of course consequences of the mirror symmetry.

The Picard-Lefshetz theory we have used above controles how
branes are created when a brane passes through a critical point
of the superpotential. Under mirror symmetry, this is mapped to
the operation called {\it mutation of bundles} \cite{rudakov}.
For this and other things, including identification of the 
mirror of other bundles, see \cite{HIV}.
The same subject is also studied in \cite{Seidel}.

\subsection{A-Branes on $X$
$\longleftrightarrow$ B-Branes of LG}

Let us consider the general toric manifold $X$.
One can construct a boundary interaction in the linear sigma model
that includes a boundary analog of F-term
which depends holomorphically
on parameters $s_1,...,s_N$ obeying $\sum_{i=1}^NQ_i^as_i=t^a$.
We denote $s_j=c_j-ia_j$.
In the sigma model limit $e_a\to\infty$, the boundary interaction reduces
to the one for the D-brane wrapped on the torus $T^{N-k}_c$ defined by
$|\phi_i|^2=c_i$ with the Wilson line
\begin{equation}
A_a=\sum_{i=1}^N\left[\, a_i\dd\varphi_i-\theta^aM_{ab}Q_i^bc_i\dd\varphi_i
\, \right],\label{wilson}
\end{equation}
where $\varphi_i=\arg\phi_i$ and $M_{ab}$ is the inverse matrix
of $\sum_{i=1}^NQ_i^aQ_i^bc_i$.
(\ref{wilson}) is well-defined on $X$ if and only if the constraints
$\sum_{i=1}^NQ_i^as_i=t^a$ are satisfied.
Dualizing the boundary theory,
we find that the mirror of $(T_c,A_a)$ is a D0-brane in the LG model
located at $\e^{-Y_i}=\e^{-s_i}$.
Supersymmetry is not broken if this point is a critical
point of (\ref{Wmir}). This gives an unexpected constraint
on $T_c$ and $A_a$.
For example, consider $X=\CP^1$ where the mirror is
the ${\cal N}=2$ sine-Gordon model $W=\e^{-Y}+\e^{-t+Y}$
There are two critical points $\e^{-Y}=\pm\e^{-t/2}$.
The corresponding branes are wrapped on the equator $T_{r/2}$
and has Wilson line $0$ or $\pi$.
(This constraint seems to correspond to the absence
of an anomaly in the nilpotency $(Q_A)^2=0$ of
the open string supercharge $Q_A$
in the definition of the Floer homology group \cite{FOOO}.
We thank K.~Fukaya for discussion on this.)

One can use this mirror symmetry to solve
open topological sigma model for $(T_c,A_a)$
in terms of open topological LG model for the corresponding D0-brane.
Let $(\theta)^n:=\overline{\theta}^1\cdots\overline{\theta}^n$ be the
chiral ring element of the boundary LG model.
This corresponds to the volume element of $T_c$.
The cylinder amplitude with
$(\theta)^n$ insertion at both boundaries is given by
$\det\partial_i\partial_jW$ at the D0-brane location $p$.
The disk amplitude with boundary $(\theta)^n$ insertion
and a bulk chiral ring element $O$ inserted in the interior
is the value of $O$ at $p$.
It is a simple exercise to show, for $X=\CP^1$,
this gives the correct number of
holomorphic disks with the boundary lying in the equator $T_{r/2}$.

One may consider promoting the paramaters $s_i$ to chiral
superfields on the boundary. 
This will give the mirror D-brane a higher dimension.
However, the superpotential has to be constant on the brane for $Q_B$
to be preserved. This gives a severe constraint on the way
$s_i$ are promoted to fields $S_i$. 
One possibility, motivated by \cite{AV},
is $S_i=Z+\tilde{s}_i$
where $Z$ is a field and $\tilde{s}_i$ are parameters,
which is possible only for $b_1^a=0$ (by the condition
$\sum_{i=1}^NQ_i^aS_i=t^a$), i.e. when $X$ is Calabi-Yau.
Then $W\!=$ constant if $\sum_{i=1}^N\e^{-\tilde{s}_i}=0$.
This condition corresponds, in an asymptotic region,
to the smoothness of the D-brane in the sigma model side.
In fact, the D-brane is wrapped on a special Lagrangian submanifold
of $X$ (the one-loop running of $c_i$ can
be absorbed by the shift of $Z$).
There are several asymptotic regions,
and different regions correspond to different class of submanifolds.
This picture is discussed in \cite{AV},
where the space-time superpotential is also computed.

\section{3d Mirror Symmetry}\label{sec:3d}

There is a class of dualities in
$2+1$ dimensional supersymmetric gauge theories
which is also known as mirror symmetry \cite{ISmir}.
It exchanges the Coulomb branch of one theory and the Higgs branch
of the dual.
Mirror symmetry of ${\cal N}=2$ abelain gauge theories
in $2+1$ dimensions is actually related \cite{AHKT} to
the $1+1$ dimensional mirror symmetry.

We consider the following mirror pair of ${\cal N}=2$
Abelian Chern-Simons gauge theories \cite{nd},
obtained by RG flow from the
${\cal N}=4$ Abelian mirror pair \cite{dhooy}.

\noindent
{\bf Model A:}\ $U(1)^k$ gauge theory with chiral multiplets $\Phi_i$ 
($i\!=\!1,...,N$) of charge $Q_i^a$ ($a\!=\!1,...,k$),
with the Chern-Simons couplings
$k^{ab}={1\over 2}\sum_{i=1}^NQ^a_{i}Q^b_{i}$. 
Gauge coupling constants $e_a$,
FI parameters $\zeta^a$,
real masses $m_i$.

\newcommand{\hPhi}{\widehat{\Phi}}
\newcommand{\hz}{\widehat{\zeta}}
\newcommand{\hm}{\widehat{m}}
\newcommand{\hQ}{\widehat{Q}}
\newcommand{\hgamma}{\widehat{\gamma}}

\noindent
{\bf Model B:}\ $U(1)^{N-k}$ gauge theory with chiral multiplets
$\hPhi_i$ ($i\!=\!1,...,N$) of charge $\hQ_i^p$ ($p\!=\!1,...,N-k$),
with the Chern-Simons couplings
$\widehat{k}^{pq}=-{1\over 2}\sum_{i=1}^N\hQ_i^p\hQ_{i}^q$. 
Gauge coupling constants
$\widehat{e}_p$,
FI parameters $\hz^{p}$, real masses $\hm_i$.

\noindent
The two sets of charges obey
$\sum_{i=1}^NQ^a_{i}\hQ_i^p=0$ ($\forall a$, $\forall p$).
The mass and the FI parameters are related by the mirror map
\begin{equation}
\begin{array}{l}
\zeta^a-{1\over 2}\sum_{i=1}^N Q_i^am_i=\sum_{i=1}^NQ_i^a\hm_i,
\\[0.2cm]
-\sum_{i=1}^N\hQ_i^pm_i=\hz^p+{1\over 2}\sum_{i=1}^N\hQ_i^p\hm_i.
\end{array}
\label{mm}
\end{equation}

We compactify this mirror pair on the circle of radius $R$.
First consider Model A. We do this so that we obtain $1+1$ dimensional
non-linear sigma model on the toric manifold $X$ at energies below the
compactification scale. This in particular requires the FI parameter
to depend on $R$ as $\zeta^a=(b_1^a\log({1\over 2\pi R\Lambda})+r^a)/2\pi R$
where the scale $\Lambda$ and the parameter $r^a$ are fixed
in the continuum limit $R\to 0$. The masses $m_i$
are of course set equal to zero.

In Model B compactification, we treat it as the $1+1$ dimensional
gauge theory with infinitely many Kaluza-Klein matter fields.
The parameters must be fixed by the mirror map (\ref{mm}).
This requires the masses to depend on $R$ as
$\hm_i=(\log({1\over 2\pi R\Lambda})+r_i)/2\pi R$ where $r_i$ solves
$\sum_{i=1}^NQ_i^ar_i=r^a$. Also, we have
$\hz^p=-\sum_{i=1}^N\hQ_i^p\hm_i/2$.
By the mirror map of the fields, it is also appropriate to rescale
the fieldstrength as $\widehat{\Sigma}_p=\Theta_p/2\pi R$
(then $\Theta_p$ has period $2\pi i$).
In terms of this rescaled field, the tree level superpotential
coming from the CS and FI couplings is
$\widetilde{W}_{\rm CS,FI}
={1\over 8\pi R}\sum_{i,p,q}\hQ_i^p\hQ_i^q\Theta_p\Theta_q
+{1\over 2}\sum_{i,p}\hQ_i^p\hm_i\Theta_p$.
The superpotential receives quantum correction ${\Delta}\tW$
only at the one loop level.
Here we have to sum over the contributions of all the charged KK modes
of mass $\widehat{\Sigma}_j+in/R$ ($j\!=\!1,...,N$, $n\in \Z$)
where $\widehat{\Sigma}_i
:=\sum_{p=1}^{N-k}\hQ_i^p\widehat{\Sigma}_p+\hm_i$.
This is done as
\begin{equation}
{\partial\Delta\tW\over \partial\widehat{\Sigma}_p}=
-\sum_{i,n}\hQ_i^p\log(\widehat{\Sigma}_i+in/R)
=-\sum_{i=1}^N\hQ_i^p\log(\e^{\pi R\widehat{\Sigma}_i}
-\e^{-\pi R\widehat{\Sigma}_i}),
\end{equation}
where we have used the zeta function regulariation that is implicit in the
mirror map (\ref{mm}).
Because of the $R$ dependence of $\hm_i$,
$\pi R\widehat{\Sigma}_i$ is divergent
and one can expand the logarithm as
$\pi R\widehat{\Sigma}_i-\e^{-2\pi R\widehat{\Sigma}_i}+\cdots$
where the dots are the higher powers of $\e^{-2\pi R\widehat{\Sigma}_i}$.
In terms of the rescaled variables $\Theta_p$,
these dots vanish in the limit $R\to 0$, and we obtain
\begin{equation}
\Delta \tW=-{1\over 8\pi R}\sum_{i,p,q}\hQ_i^p\hQ_i^q\Theta_p\Theta_q
-{1\over 2}\sum_{i,p}\hQ_i^p\hm_i\Theta_p
-\Lambda\sum_{i=1}^N\e^{\hQ_i\cdot \Theta_p-r_i}.
\end{equation}
The linear and quadratic terms in $\Theta$ cancell the tree-level terms
$\tW_{\rm CS,FI}$ and the total is
$\tW=\tW_{\rm CS,FI}+\Delta\tW=
-\Lambda\sum_{i=1}^N\e^{\hQ_i\cdot \Theta_p-r_i}$.
This is nothing but the superpotential
(\ref{Wmir}) of the $1+1$ dimensional mirror of the toric sigma model.

This analysis of Model B compactification
is valid if the ratio $\hgamma=$ (3d gauge
coupling)$^2/$(compacification scale) is much smaller than $1$.
However, 3d mirror symmetry is originally regarded as an infra-red duality,
which would imply the equivalence of the compactified theories
only for $\hgamma\gg 1$.
Thus, the compactification does not precisely relate 3d mirror symmetry
as IR duality and 2d mirror symmetry of \S\ref{sec:proof}.
(Of course, superpotential is independent of $\hgamma$
and the compactification does relate such (twisted) F-term
information of the theories.)
There is actually a proposal \cite{KS} that after an $\widehat{e}_p$
dependent modification of Model A it is equivalent to Model B
at energies smaller than $e_a$ but arbitrary compared to $\widehat{e}_p$.
Compactification of the modified Model A yields
the sigma model on the squashed toric manifold $X'$ of \S\ref{sec:2dBH}
where $\hgamma_p$ plays the role of the squashing parameter $k_p$.
On the other hand, analysis of superpotential of Model B compactification
is same as above. Moreover, in the regime $\hgamma\ll 1$ one can also
analyze the K\"ahler potential, and this agrees with the K\"ahler potential
(\ref{KW}) of the 2d mirror.
This can be regarded as support for the 3d mirror symmetry
beyond IR duality.

\section*{Acknowledgement}

I would like to thank the organizers of Strings 2001 for a stimulating
conference. Food was especially good!
I am grateful to M.~Aganagic, A.~Iqbal, A.~Kapustin, A.~Karch,
D.~Tong and C.~Vafa for collaborations.
This work was supported in part by NSF-DMS 9709694.

\bibliographystyle{amsalpha}

\end{document}